# Thermal radiation of conducting nanoparticles

Yu.V.Martynenko, L.I.Ognev

*Nuclear Fusion Institute, Russian Research Center "Kurchatov Insitute",
123182, Moscow, Kurchatov sq., 1*

Abstract

The thermal radiation of small conducting particles was investigated in the region where the Stephan-Boltzmann law is not valid and strongly overestimates radiation losses. The new criterion for the particle size, at which black body radiation law fails, was formulated. The critical radius $r_c$ is expressed through a combination of temperature $T$ and particle conductivity $\sigma$: thus $r_c = c(\hbar/2\pi\sigma kT)^{1/2}$. The approach is based on the magnetic particle polarization, which is valid until very small sizes (cluster size) where due to drop of particle conductivity the electric polarization prevails over the magnetic one. It was also shown that the radiation power of clusters, estimated on the basis of the experimental data, is lower than that given by the Stephan-Boltzmann law.

In the last years the amount of research in the field of nanotechnology, nanomaterials and nanosystems tremendously increased [1]. One of the most important trends in the nanotechnology is obtaining and using of nanopowders, which can be used in metallurgy, microelectronics, and medicine and food industry. Production and processing of nanopowders are frequently dealt with heating of nanoparticles to high temperatures when intensive thermal radiation can be expected.

The problem is close related to dust plasma [2], and to the role of dust in fusion devices [3], where micro- and nano- particles are heated to high temperatures and radiation energy losses can play decisive role in the energy balance of the particles. The effect of metal clusters on emission spectra was discussed in [4].

The black body radiation cannot be used for the calculation of nanoparticle energy balance, because for the particles smaller than radiation wavelength the Stephan-Boltzmann formula gives far overestimated results.

For the calculation of a small body energy loss the Kirchhoff's law of thermal radiation can be used [5]. Thermal radiation of the body with the temperature $T$ in the interval of angular frequencies $d\omega$ equals

$$dI(\omega) = 4\pi c \sigma(\omega) \frac{\hbar\omega^3}{4\pi^3 c^3 (e^{\frac{\hbar\omega}{kT}} - 1)} d\omega, \tag{1}$$

where $\sigma(\omega)$ is the effective cross section for frequency $\omega$, $c$ is the light velocity, $T$ is the temperature, $k$ is the Boltzmann constant.

According to [5], the radiation absorption cross section for the sphere with the volume $V$ can be expressed through electric $\alpha''_e$ and magnetic $\alpha''_m$ polarization per unit volume,

$$\sigma(\omega) = \frac{4\pi\omega}{c}(\alpha''_e + \alpha''_m) \cdot V \tag{2}$$

For conducting particles the magnetic polarization prevails with the exception of very small sizes (see below). Let us consider magnetic $\alpha''_m$ polarization part of absorption cross section. The radiation of atomic clusters will be discussed later.

The theory of scattering and absorption of electromagnetic radiation on small particles developed by G. Mie [7] is not possible to give sufficiently common dependencies for the cross section on parameters of particles. In the case of conducting spheres the approximate approach can be used which is based on the account of radiation penetration depth in the matter [5]. In the frame of this approach the magnetic polarization of a sphere with radius $r_0$ depends on dimensionless parameter $(r_0/\delta)$, where $\delta = c/(2\pi\sigma\omega)^{1/2}$ is the radiation penetration depth into conductor as

$$\alpha''_m = -\frac{9\delta^2}{16\pi r_0^2}\left[1 - \frac{r_0}{\delta}\frac{sh(2r_0/\delta) - \sin(2r_0/\delta)}{ch(2r_0/\delta) - \cos(2r_0/\delta)}\right] \tag{3}$$

For small radiuses $(r_0/\delta \ll 1$, low frequencies) the following approximation is possible

$$\alpha_m^" = \frac{1}{20\pi}\left(\frac{r_0^2}{\delta^2}\right) = \frac{r_0^2 \sigma \omega}{10c^2}. \qquad (4)$$

In this case ($r_0/\delta \ll 1$) absorption cross-section scales as $\sigma_m \sim \omega^2$ and therefore thermal radiation of small particles has maximum shifted to frequencies higher than that given by Wien law.

$$\lambda_{max}(r/\delta \ll 1) = 3B/5T, \qquad (5)$$

where $B = 0.29$ cm·K is the Wien constant.

For large radiuses of particles ($r_0/\delta \gg 1$, large frequencies) the approximation

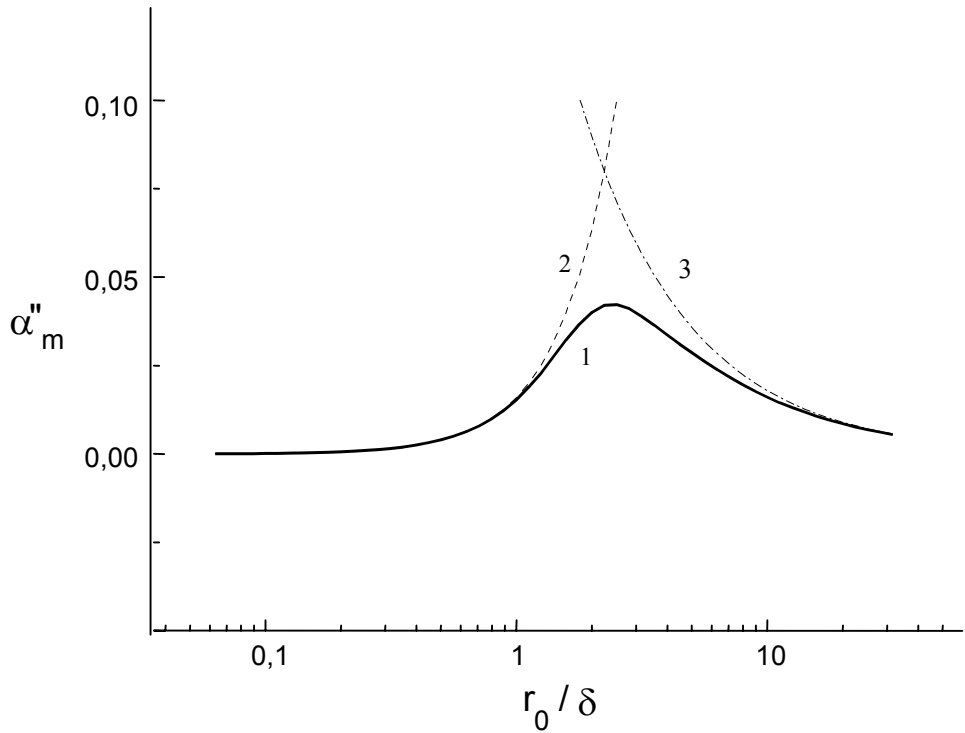

Fig. 1. Dependence of imaginary part of magnetic polarization per unit volume $\alpha_m^"$ on dimensionless parameter ($r_0/\delta$) for exact formula (curve 1) and for approximations at small frequencies ($r_0/\delta \ll 1$, curve 2) and high frequencies ($r_0/\delta \gg 1$, curve 3).

$$\alpha_m^" = \frac{9\delta}{16\pi r_0} = \frac{9c}{16\pi r_0 \sqrt{2\pi\sigma\omega}} \qquad (6)$$

can be used. Here σ is the static conductivity of a substance.

Dependencies (3), (4) and (6) are shown in the Fig. 1. Substitution of the expression (6) for high energies into the integral for radiation (1) results in disappearance of the radius of a particle $r_0$ from final formula. It means transition to the black body thermal radiation.

Introducing dimensionless parameter $p$, which takes into account conductivity of a substance, particle radius and its temperature,

$$p = \frac{r_0}{c}\sqrt{\frac{2\pi\sigma kT}{\hbar}}, \tag{7}$$

one can reduce the expression (1) for sphere thermal radiation to

$$I = \frac{16}{3} \cdot \frac{T^5 r_0^3}{c^3 \hbar^4} \cdot J(p). \tag{8}$$

Dependence of the integral $J(p)$ on dimensionless parameter $p$,

$$J(p) = \int_0^\infty \alpha_m^{''}(p\sqrt{x}) \frac{x^4 dx}{e^x - 1}, \tag{9}$$

is shown in the Fig. 2. In the expression (9) dimensionless variable $x = \hbar\omega/kT$ was used. At $p \gg 1$ $J(p) \sim 1/p$, that corresponds the case of high temperatures or large radiuses of the particles, whereas at $p \ll 1$ quadratic asymptotic $J(p) \sim p^2$ is valid. In the last case asymptotic shows high power dependence of normalized cross section $\sigma(\omega)/(\pi r_0^2)$ on radiuses, which follows also from the Mie's theory [7].

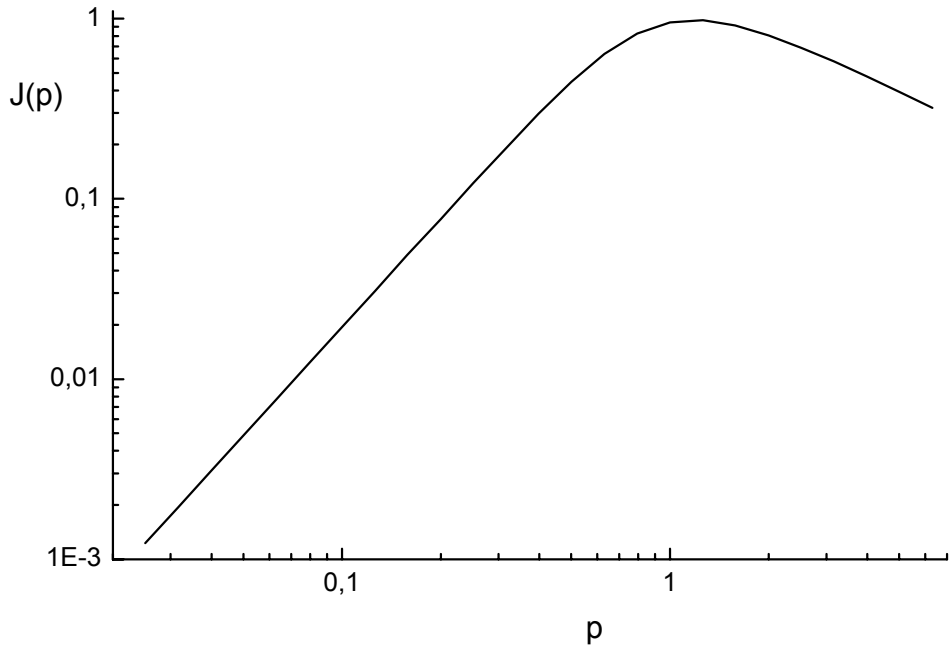

Fig. 2. Dependence of the integral *J(p)* on dimensionless parameter *p*.

The expression for the thermal radiation intensity normalized by black body radiation $I_b$ is

$$\frac{I}{I_b} = \frac{80kT}{\pi^3 c\hbar} \cdot r_0 J(p) \tag{10}$$

Due to above mentioned asymptotic $J(p) \sim 1/p$ in the case of large particles the dependence of normalized intensity on radius disappears. The temperature dependence of $I/I_b$ at large $r$ ($I/I_b \sim T^{1/2}$) can be accounted for not perfect accuracy of the model for absorption of radiation by conducting balls at high temperatures.

The dependences of normalized radiant emittance ("blackness") $I/I_b$ on particle radius $r_0$ at temperatures $T=1773$ °K, 1273 °K and 773 °K for copper particles with account for temperature dependence of conductivity of the material [8] are shown in the Fig.3.

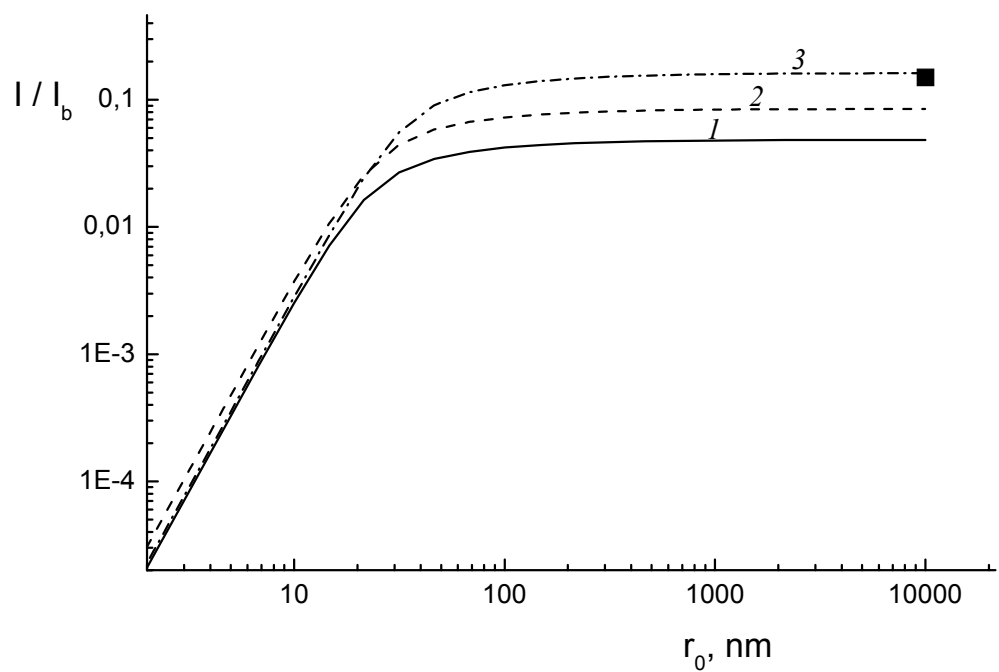

Fig. 3    Normalized radiant emittance of copper particles at temperatures 1773 °K (3) , 1273 °K (2) and 773 °K (1).

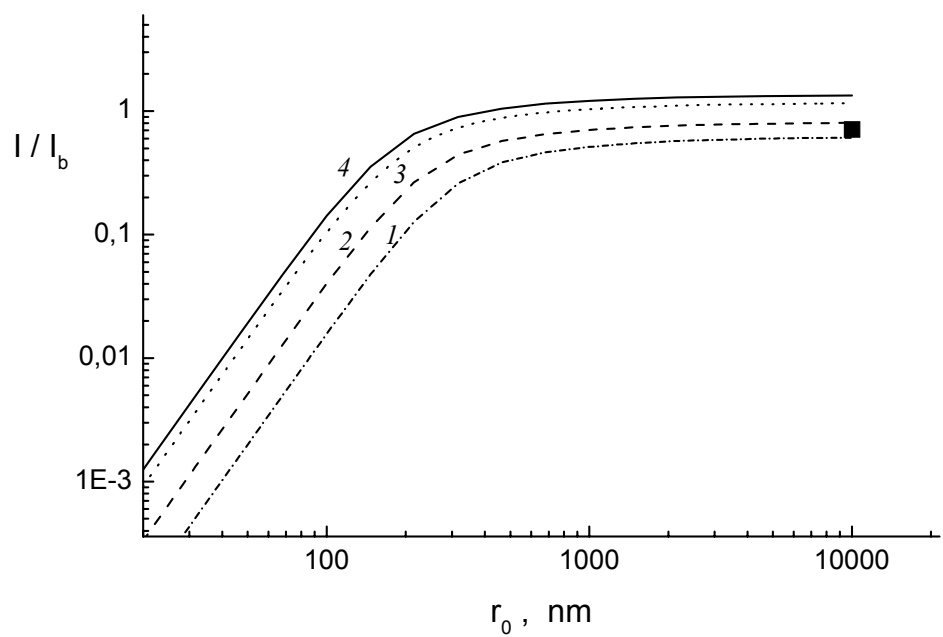

Fig. 4. Normalized radiant emittance of graphite particles at temperatures 2773 °K (4), 2273 °K (3), 1273 °K (2) and 773 °K (1).

The similar results for micro particles of graphite at temperatures $T$= 2773 °K, 2273 °K, 1273 °K and 773 °K are shown in the Fig. 4.

The dependence of radiant emittance normalized by black body radiant emittance, $I/I_b$ ("blackness") on spherical particles dimensions looks like qubic dependence of normalized cross section of absorption on particle radius and leads to a constant value for large radius particles. For molten cooper at temperatures 1100 – 1300 °C "blackness" equals 0.13-0.15 [8], that is in a good agreement with calculations for particles with the radius greater than 1 μm (see Fig. 3).

The similar results for micro particles of graphite at temperatures $T$= 2773 °K, 2273 °K, 1273 °K and 773 °K are shown in the Fig. 4. For graphite at temperature 500 °C "blackness" equals to 0.71 [9] that agrees with the results shown in the Fig. 4. Overestimated results of graphite radiant emittance at elevated temperatures can be explained by inaccuracy of the model, which is valid for frequency $\omega/(2\pi) \ll \sigma$. Normalized cross-section of absorption at spherical particles for high frequencies, corresponding to asymptotic (5), can be reduced to $\sigma(\omega)/(\pi r_0^2) = 3[\omega/(2\pi\sigma)]^{1/2}$. For more precise model of radiation absorption by conducting substance in the asymptotic also a terms of an expansion of the order of $\omega/(2\pi\sigma)$ should exist [7]. It is also worth noting that in practically important case of metal carbides the criterion is of the validity of the approach is well satisfied.

For both examples of copper and graphite considerable decrease of radiant emittance compared with Stephan-Boltzmann black body law for small radiuses takes place at $p \leq 1$. So the dimensionless parameter $p=(r_0/c)\cdot(2\pi\sigma kT/\hbar)^{1/2}$ characterizes radiant emittance of small conducting particles depending on their dimension and conductivity.

Let us compare radiation absorption cross section calculated with magnetic $\alpha''_m$ polarization at small radiuses ($r_0/\delta \ll 1$, see (4)),

$$\sigma_m = (4\pi\omega/c)\cdot(4\pi r^3/3)\cdot\alpha_m'' = (16\pi^2/30)\cdot\omega^2\sigma r^5/c^3 \tag{11}$$

and cluster absorption cross section obtained experimentally and given in [6]. One should keep in mind that small particles have not crystalline structure and its conductivity is $\sigma \approx 10^{15}$ s$^{-1}$. Really

$$\sigma = (ne^2/mv_e) \cdot \lambda_{path} \qquad (12)$$

where $\lambda_{path}$ is the electron free path, $v_e$ is the electron velocity in solid, $n$ is the electron density in cluster. For crystals $\lambda_{path}$ is determined by electron scattering on phonons and on defects, whereas for non-crystalline matter $\lambda_{path} \approx a$, where $a$ is the distance between the atoms. Estimation of $\sigma$ at $\lambda_{path} = a$ gives $\sigma \approx 10^{15}$ s$^{-1}$.

Radiation absorption for small clusters is not clear now [6]. It may be determined by radiation transition between spectral bands, or by free electron plasma oscillation excited by electric field [6]. But in any case absorption of small clusters is determined by electric polarization.

In the Fig. 5 the cross sections $\sigma_m$ (10) for Ag cluster are shown as functions of $r$ at frequency $\hbar\omega = \hbar\omega_0 = 4$ eV where cross sections have maximum value. Curve 1 corresponds to magnetic polarization model corrected for small cluster radiuses, curve 3 corresponds to magnetic polarization model with cluster size conductivity $\sigma \approx 10^{15}$ s$^{-1}$ and curve 2 corresponds to fit to experimental data for cross section of Ag particles [10] with radius dependence of electric polarization model as $\sigma_e \sim r^3$ [6].

For $r < 18$ nm electric polarization prevails $\sigma_e > \sigma_m$. It means that at $r < 18$ nm absorption is determined by electric polarization, $\sigma_{abs} \approx \sigma_e$, and for $r$ 18 nm $\sigma_{abs} \approx \sigma_m$. If transition amorphous- crystalline structure of clusters occurs at radius $r_{tr} < 18$ nm conductivity $\sigma$ grows abruptly near $r_{tr}$, absorption cross section $\sigma_m$ grows also abruptly at $r_{tr}$ and transition from electric polarization absorption mechanism to that of magnetic polarization occurs at $r \approx r_{tr}$. This transition matching $\sigma_m$ and $\sigma_e$ is shown on curve 1 on Fig. 5 for the case of amorphous - crystalline structure transition at radius $r_{tr} = 7$ nm

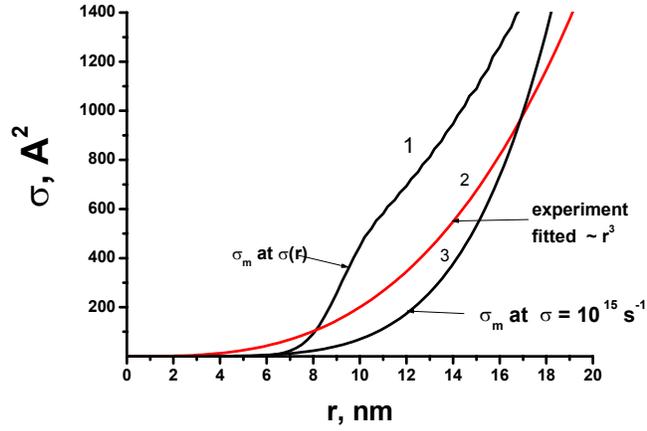

Fig.5. Absorption cross-sections $\sigma_m$ calculated by (10) matched to $\sigma = 10^{15}$ s$^{-1}$ (curve 1) and with $\sigma(r)$ and $\sigma_e$ which is Ag clusters cross section [10] extrapolated in accordance $\sigma_e \sim r^3$ [6] (curve 2). Curve 3 corresponds to magnetic model with reduced conductivity as in the amorphous media.

In the example shown on Fig. 5 the maximum value for the electric polarization cross section $\sigma_e$ at $\omega_0$ was taken. Apart from the maximum the electric polarization would be not so high and for particle size higher than about 10 nm the magnetic polarization prevails. Thus the theory of thermal radiation based on the magnetic polarization of conducting particles that was developed is valid for $r > 10$ nm and describes transitional region from small particles thermal radiation to that of the black body.

Note that cluster absorption cross sections have resonant shape (sometimes several resonances) with the resonant frequency higher than the maximum of black body radiation at reachable temperatures and closer to the visible light spectrum. Therefore hot clusters, being a good light source, may have still low radiation power as compared with black body particles.

Let us compare the average cluster radiation powers per unit mass, $P_{rad}$, calculated in [6] by resonant-like cross section $\sigma_{abs}$ for Ag, Li and K clusters and the black sphere radiation power per unit mass, $P_b$.

$$P_b = 3\pi r^2 \sigma_{SB} T^4 / 4\pi r^3 \rho, \qquad (13)$$

where $\sigma_{SB}$ is the Stephan-Boltzmann constant, $\rho$ is the density of the particle. In the Table I there are $P_{rad}$ and $P_b$ for Ag, Li and K clusters. The calculated [6] values of $P_{rad}$ are independent of $r$, whereas black sphere radiation $P_b \sim r^{-1}$ and for considered clusters the black sphere radiation power per unit mass $P_b$ is much higher than that for clusters.

Table I

| Cluster | 3500 K | | 4000 K | |
|---|---|---|---|---|
| | $P_{rad}$, $10^7$ W/g | $P_b$, $10^7$ W/g | $P_{rad}$, $10^7$ W/g | $P_b$, $10^7$ W/g |
| $Ag_9$ | 1.6 | 19,5 | 3.5 | 32 |
| $Ag_{21}$ | 1.6 | 14.7 | 3.5 | 24 |
| $Li_{139}$ | 4.9 | 146 | 10 | 298 |
| $Li_{270}$ | 4.9 | 117 | 10 | 239 |
| $Li_{440}$ | 4.9 | 99 | 10 | 202 |
| $Li_{820}$ | 4.9 | 81 | 10 | 165 |
| $Li_{1500}$ | 4.9 | 66 | 10 | 135 |
| $K_9$ | 8.6 | 266 | 17 | 457 |
| $K_{21}$ | 8.6 | 201 | 17 | 345 |
| $K_{500}$ | 8.6 | 70 | 17 | 120 |
| $K_{900}$ | 8.6 | 58 | 17 | 99 |

Therefore hot cluster thermal radiation is much less then that of black body. However hot clusters are more bright than black body, because their radiation spectrum maximum is closer to the most visible frequencies.

In the case of dielectric particles it is not possible to develop rather simple dependence of radiation absorption cross-section on the radius of the spherical particle. But G. Mie theory predicts slow variation of absorption cross-section for wavelengths $2\pi\lambda(n-1)/r_o \leq 6$, where $n$ is the refraction index. Since maximum of thermal radiation spectrum due to Wien's displacement law [8] is at the wavelength $\lambda_{max} = B/T$, the considerable decrease of radiant emittance with particle size decrease is expected for particle dimensions $r_0 \leq (n-1)B/T$.

CONCLUSION

The radiant emitance of small conducting particle is lower than that given by Stefan-Boltzmann law at size below a critical size and drops with particle dimension decrease. The universal criterion for the particle size, at which black body radiation law fails, was formulated. The critical radius $r_c$ is expressed through a combination of temperature $T$ and particle conductivity $\sigma$, thus $r_c = c(\hbar/2\pi\sigma kT)^{1/2}$. The radiation of conducting particles is determined by magnetic polarization at $\omega/(2\pi) < \sigma$. At very small sizes (cluster size) conductivity of particles drops and the electric polarization prevails over magnetic one. Radiation power of clusters, estimated on the basis of experimental data, is also lower than that given by Stephan-Boltzmann law.

The work is supported by RFBR grant 04-02-08180 obr-a.

REFERANCES